\documentclass[twocolumn]{aastex702}
\usepackage{amsmath}
\usepackage{subfiles}

\newcommand{\Euclid}{\textit{Euclid}}

\shorttitle{Agentic Strong-Lens Modeling in Euclid Q1}
\shortauthors{Cao et al.}

\begin{document}

\title{Toward Automated Strong-Lens Modeling with Agentic Workflows: An Application to \Euclid\ Q1 Lenses}

\correspondingauthor{Xiaoyue Cao and Ran Li}

\author[orcid=0000-0003-4988-9296]{Xiaoyue Cao}
\affiliation{Institute for Astrophysics, School of Physics, Zhengzhou University, Zhengzhou 450001, China}
\email[show]{xycao@zzu.edu.cn}

\author[orcid=0000-0003-3899-0612]{Ran Li}
\affiliation{School of Physics and Astronomy, Beijing Normal University, Beijing 100875, China}
\email[show]{liran@bnu.edu.cn}

\author[orcid=0009-0009-5829-9053]{Yunlong Yin}
\affiliation{National Astronomical Observatories, Chinese Academy of Sciences, 20A Datun Road, Chaoyang District, Beijing 100012, China}
\affiliation{School of Astronomy and Space Science, University of Chinese Academy of Sciences, Beijing 100049, China}
\email{yinyl@bao.ac.cn}

\author[orcid=0000-0002-3490-4089]{Rui Li}
\affiliation{Institute for Astrophysics, School of Physics, Zhengzhou University, Zhengzhou 450001, China}
\email{liruiww@zzu.edu.cn}

\author{Liang Gao}
\affiliation{Institute for Astrophysics, School of Physics, Zhengzhou University, Zhengzhou 450001, China}
\affiliation{School of Physics and Astronomy, Beijing Normal University, Beijing 100875, China}
\email{lgao@bnu.edu.cn}

\begin{abstract}
Ongoing and forthcoming wide-field surveys will discover $\mathcal{O}(10^5)$
galaxy--galaxy strong lenses, shifting the main bottleneck from detection to
reliable physical modeling. Conventional forward-modeling pipelines still rely
on case-by-case expert judgment and manual tuning to mask contaminants,
initialize priors, diagnose unphysical fits, and reconfigure failed runs. We
present Lens-Orchestra, an automated workflow coupling a team of multimodal
large language model agents to multi-stage physical modeling with elliptical
power-law mass profiles, external shear, and pixelized source reconstructions.
Launched via natural-language instructions, it modeled 185 \Euclid\ Quick Data
Release (Q1) grade-A lens candidates with published mass models without human
intervention, completing 149 fits. Independent human evaluation confirmed 140
usable models---including 126 grade-I fits with noise-level residuals, with only
$6.0\%$ of completed fits unacceptable---and endorsed all 36 halted runs. Across
the 140 usable systems, effective Einstein radii match released Q1 models
($+0.12\%$ median offset; $2.73\%$ typical scatter), while hierarchical
inference on the 126 grade-I imaging-only fits yields a mean mass-density slope
of $2.002 \pm 0.009$ (intrinsic scatter $0.074^{+0.007}_{-0.006}$). A dedicated
residual-screening agent also guided the selection of two candidate low-mass
perturbers for parametric and free-form substructure modeling; although neither
was confirmed, this end-to-end search illustrates how automated screening
focuses computationally intensive substructure analyses on a few promising
systems. Coupling
multimodal agents to explicit forward modeling thus enables accurate,
survey-scale lens analysis from macro mass profiles to small-scale dark matter
structure.
\end{abstract}


\section{Introduction}\label{sec:intro}
Galaxy--galaxy strong gravitational lensing maps the projected mass distribution
of deflector galaxies through the multiple images and extended arcs of
background sources. Combined with stellar-population or dynamical data, strong
lensing separates stellar and dark matter contributions and constrains the
stellar initial mass function
\citep{TreuKoopmans2004,Treu2010,Auger2010}, while localized surface-brightness
perturbations along extended arcs probe low-mass dark subhalos and line-of-sight
halos that test cold, warm, and self-interacting dark matter models
\citep[e.g.,][]{Vegetti2010,Vegetti2012,Li2016,He2022,Vegetti2024,Nightingale2024,Cao2025,Powell2025,ShuboLi2025}.
Ongoing and forthcoming space-based surveys with \Euclid, the China Space
Station Telescope (CSST), and the Nancy Grace Roman Space Telescope will detect
$\mathcal{O}(10^5)$ strong lenses \citep{Collett2015,Cao2024,Wedig2025}---up to
three orders of magnitude more than legacy Hubble Space Telescope (HST) samples
\citep{Bolton2008SLACS,Brownstein2012BELLS,Sonnenfeld2013SL2S,Shu2016BELLSGALLERY}.
Such samples will enable statistical constraints on galaxy structure and dark
energy from $\mathcal{O}(10^3\text{--}10^4)$ lenses
\citep{SonnenfeldCautun2021,Li2024,Wu2026CSST}, and on the nature of dark matter
from thousands of detectable subhalos \citep{ORiordan2023EuclidSubstructure}.

Exploiting these datasets requires accurate pixel-level forward modeling that
reconstructs the lensed emission to the observational noise level, shifting the
main bottleneck in strong-lensing science from detection to reliable modeling.
Neural networks can rapidly estimate lens parameters and uncertainties directly
from images \citep{hezaveh2017fastcnnlensmodelling,gentile2023lemoni}, but
remain sensitive to mismatches between training simulations and real lenses
\citep{wagnercarena2021hierarchicalbnninference}. A complementary approach
retains explicit forward modeling while automating its execution: neural
networks can be used to initialize physical lens models to aid sampler
convergence
\citep{pearson2021cnnplusparametricfitting,euclidbusillo2026euclidq1lemon} or,
in the \texttt{lenstronomy}-based \texttt{dolphin} pipeline, to segment image
components before fitting \citep{BirrerAmara2018,shajib2025dolphin}, while
\texttt{AutoLens} and its Source Light and Mass (SLaM) pipelines decompose a
complex analysis into a sequence of simpler stages in which each fit initializes
the next \citep{nightingale2018autolens,etherington2022nolensleftbehind}. These
developments automate important
components of lens modeling, but not yet the full decision loop: as illustrated
by the blind application of SLaM to 59 HST lenses
\citep{etherington2022nolensleftbehind}, human experts still have to mask
unrelated sources, mark multiple-image positions to steer the sampler away from
local maxima, and inspect intermediate diagnostics to reconfigure unsuccessful
fits. Because such case-by-case manual tuning cannot scale to
$\mathcal{O}(10^5)$ lenses, survey-scale analysis requires automating the expert
decisions that surround physical forward modeling
\citep{euclidnightingale2026eclipse_c}.

Multimodal large language model (LLM) agents offer a promising way to automate
these expert decisions by combining visual inspection with the execution of
external analysis tools \citep{Yao2023,OpenAI2023}. Applied to lens modeling, an
agent can inspect survey imaging, configure and run a physical forward model,
evaluate visual and numerical diagnostics, and decide whether to refine the
configuration or flag anomalous systems for follow-up.

In this Letter, we introduce Lens-Orchestra%
\footnote{The lens-model catalog and diagnostic plots from this work are provided at \url{https://github.com/caoxiaoyue/lens-orchestra}; the full Lens-Orchestra codebase will be made publicly available in the same repository upon acceptance.},
an agent-driven forward-modeling
workflow that couples a team of multimodal agents to multi-stage physical lens
modeling in \texttt{TinyLensGPU}%
\footnote{\url{https://github.com/caoxiaoyue/TinyLensGpu}}
\citep{Cao2025TinyLens}, and evaluate it on 185
grade-A galaxy--galaxy strong-lens candidates from the \Euclid\ Quick Data
Release (Q1) \citep{euclidwalmsley2026euclidq1sldea}. Initiated through
high-level natural-language instructions, the agents automate contaminant
masking, visual prior initialization, and stage-by-stage diagnostic review and
refitting without human intervention, while recording all intermediate
diagnostics and decisions for optional human inspection.

\section{Data and Sample}\label{sec:data}
In this work, we analyze 185 grade-A galaxy--galaxy strong-lens candidates
from the \Euclid\ Quick Data Release (Q1) that have publicly released lens
mass models.\footnote{The Q1 lens catalog, imaging data, and released
modeling results can be accessed at \url{https://doi.org/10.5281/zenodo.15025832}.} Q1 covers
$63.1\,\mathrm{deg}^{2}$ across the Euclid Deep Fields North, South, and
Fornax, observed to the nominal depth of the Euclid Wide Survey
\citep{euclidaussel2026euclidq1overview}. The candidates were identified
through the Strong Lensing Discovery Engine, which combines deep-learning
searches, citizen-science inspection through Space Warps, and expert visual
vetting. Grade A denotes high-confidence candidates exhibiting clear lensing
features \citep{euclidwalmsley2026euclidq1sldea}.

For each system, we model the \Euclid\ Visible Instrument (VIS) image,
using the accompanying root-mean-square
noise map and point spread function (PSF). The input images contain
$300\times300$ pixels at a pixel scale of $0.1^{\prime\prime}$, corresponding
to a field of view of $30^{\prime\prime}\times30^{\prime\prime}$. The VIS PSF
has a typical full width at half maximum of approximately
$0.16^{\prime\prime}$, resolving the extended arcs and multiple images used
to constrain the lens models \citep{euclidwalmsley2026euclidq1sldea}.
Near-infrared imaging in the
$Y_{\rm E}$ and $J_{\rm E}$ bands is combined with
the VIS imaging to construct color images for visual inspection by the
agents. Lens modeling uses only the VIS imaging.

\section{Agent-driven Physical Lens Modeling}\label{sec:method}
\begin{figure*}[t]
    \centering
    \includegraphics[width=\textwidth]{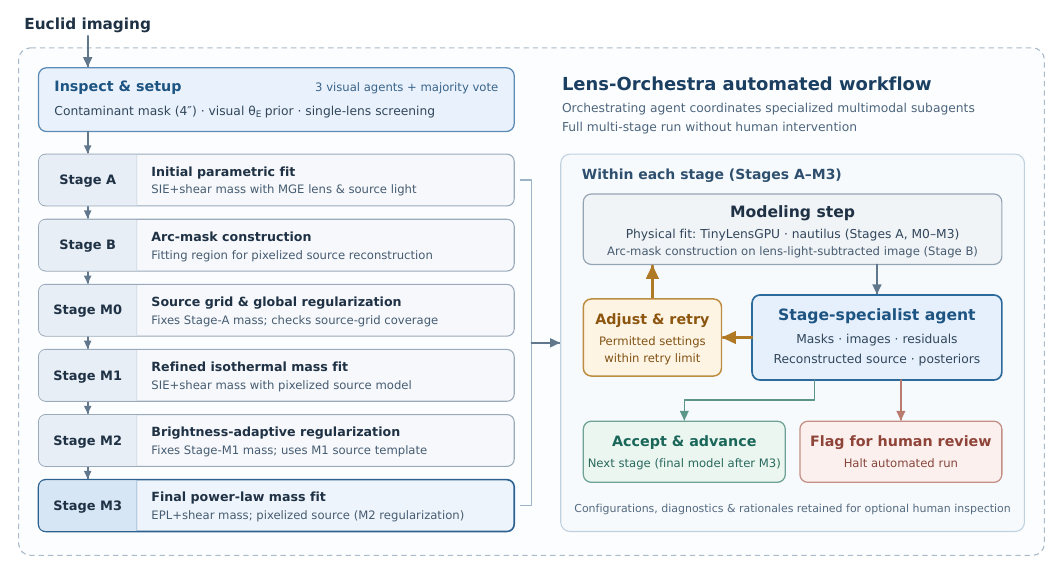}
    \caption{Schematic of the Lens-Orchestra automated workflow. An
    orchestrating agent coordinates specialized multimodal subagents across
    pre-modeling inspection and six modeling stages (Stages~A--M3; left) without
    human intervention. Three independent visual-inspection agents combine their
    assessments by majority vote to mask contaminants, estimate a visual
    Einstein-radius prior, and screen out complex or ambiguous systems for
    human review. Each stage then executes the within-stage loop (right): the modeling step
    performs a physical fit with \texttt{TinyLensGPU} and \texttt{nautilus}
    (Stages~A and M0--M3) or constructs the arc mask on the lens-light-subtracted
    image (Stage~B), after which a stage-specialist agent evaluates the masks,
    images, residuals, reconstructed source, and posteriors to either accept the
    stage and advance (delivering the final EPL+shear model after Stage~M3),
    adjust permitted settings and retry (amber loop), or halt the run and flag
    the system for human review. All configurations, diagnostics, and decision
    rationales are retained for optional human inspection.}
    \label{fig:agent-workflow}
\end{figure*}

Our workflow couples multimodal LLM agents to a prescribed multi-stage
modeling pipeline (Figure~\ref{fig:agent-workflow}). Agents configure
the analysis and evaluate its diagnostics, while \texttt{TinyLensGPU}
\citep{Cao2025TinyLens} performs physical lens modeling and parameter
inference.

\subsection{Multi-stage physical lens modeling}\label{sec:staged-inference}
We begin with a joint fit to the mass and light of the deflector and the light
of the background source. The initial mass model is a singular isothermal
ellipsoid (SIE; \citealt{Kormann1994}) with external shear, and both light
profiles are represented by multi-Gaussian expansions (MGE;
\citealt{He2024MGE}). After subtracting the fitted deflector light, we
construct an arc mask enclosing the lensed images to define the region fitted
when reconstructing the source on a pixelized grid
\citep{WarrenDye2003,Suyu2006}, and then use that reconstruction to refine the
mass model and calibrate the source regularization. The final fit
allows the mass profile to depart from isothermal by adopting an elliptical
power law (EPL; \citealt{TessoreMetcalf2015}) with external shear. We use
\texttt{nautilus} \citep{Lange2023Nautilus} to sample nonlinear parameters,
evaluating the image $\chi^2$ likelihood for the initial parametric fit and the
Bayesian evidence for the pixelized-source fits \citep{Suyu2006} (see
Appendix~\ref{app:physical-modeling} for details). Because each step builds
directly on the preceding results, an unmasked contaminant, a missed
counterimage, or a poor intermediate fit can propagate downstream and bias
the final mass model. In conventional pipelines, preventing such failures
requires human modelers to inspect intermediate fits and adjust settings step by
step; the next subsection describes how Lens-Orchestra automates these
decisions.

\begin{figure*}[tp]
    \centering
    \includegraphics[width=\textwidth]{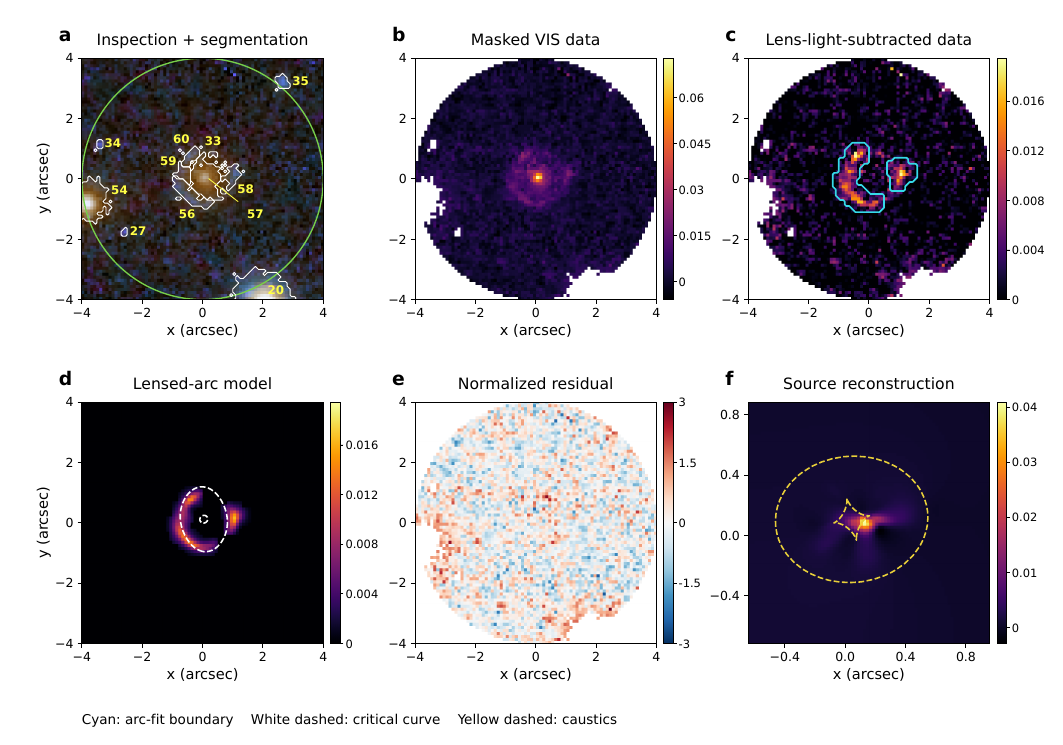}
    \caption{Multi-stage lens modeling for the grade-I system
    \texttt{102019595\_NEG630246532498667881}.
    (a) $Y_{\rm E}$/$J_{\rm E}$/VIS color composite with VIS segmentation
    contours and object IDs overlaid; segments 20, 27, 34, 35, and 54 were
    identified and masked as contaminants by the visual-inspection agents
    within the $4^{\prime\prime}$-radius modeling aperture (green circle).
    (b) VIS image after contaminant masking (masked regions shown in white).
    (c) VIS image after subtracting the Stage-A MGE lens-light model, with cyan
    contours outlining the Stage-B arc mask that defines the fitting region for
    the subsequent pixelized-source fits (Stages M0--M3).
    (d) Lensed-source reconstruction from the final EPL+shear model (Stage M3),
    overlaid with the critical curves (white dashed lines).
    (e) Normalized image residuals across the modeling aperture.
    (f) Reconstructed source surface brightness on the $80\times80$ pixelized
    grid ($0.02^{\prime\prime}\,\mathrm{pixel}^{-1}$), with caustics shown in
    yellow.}
    \label{fig:inspection-reconstruction}
\end{figure*}

\subsection{Multimodal agent orchestration and diagnostic review}\label{sec:agent-review}
To automate these decisions, Lens-Orchestra organizes the workflow around a team
of specialized LLM agents coordinated through natural-language instructions in
Codex CLI. An orchestrating agent manages the overall progression across the
modeling stages and delegates each task to role-specific subagents: an initial
visual-inspection team (\texttt{gpt-5.6-sol}) prepares the modeling inputs, and
dedicated stage subagents (\texttt{gpt-5.6-luna}) drive and diagnose the
individual fits. Much like a human modeling team, this division of labor gives
each agent a clean context window to focus on its specific task. Each agent is
guided by a structured prompt encoding expert diagnostic criteria and common
failure modes, and adjusts only predefined modeling settings without modifying
the modeling code or underlying physical model.

Before fitting begins, three independent visual-inspection subagents examine the
$Y_{\rm E}$/$J_{\rm E}$/VIS color composite and labeled VIS segmentation
map---analogous to multi-expert visual vetting in strong lens searches---to
identify contaminating objects and assess whether the system is suitable for a
single-deflector model. To guide the initial parametric fit, they also visually
estimate the Einstein radius from the approximate angular radius of the lensed
arcs or multiple images around the deflector center. A synthesis subagent
combines their independent assessments under a two-of-three majority rule for
both the single-lens screening decision and each contaminant segment. Selected contaminants are
masked within the $4^{\prime\prime}$-radius modeling aperture, while systems
showing complex multi-deflector structure or ambiguous lensing morphology are
routed directly to human review. This visual estimate is used to construct a
mixture prior for the initial Einstein radius, combining a Gaussian centered on
the estimated value with a broad uniform distribution. If the visual estimate is
inaccurate, the uniform component still spans the true parameter range, allowing
the sampler to recover the correct solution (see
Appendix~\ref{app:physical-modeling} for details).

At each modeling stage, a dedicated subagent examines the fitted model and its
associated diagnostic outputs. It then decides whether to advance, repeat the
fit with a targeted adjustment within the retry limit, or halt automated
modeling and flag the system for human review
(Appendix~\ref{app:physical-modeling}). The stages proceed as follows:
\begin{itemize}
\item \textbf{Stage A: Initial mass and light fit.} The joint fit combines an
SIE+shear mass model with MGE representations of the deflector and source light,
yielding initial mass parameters and a deflector-light model for subtraction.
The agent checks whether the fit reproduces the primary multiple-image
geometry and assesses whether deflector-light subtraction removes or distorts
any arc or counterimage.

\item \textbf{Stage B: Arc mask.} From the lens-light-subtracted image, the
pipeline constructs the arc mask that defines the fitting region for the
pixelized-source stages (Stages M0--M3). If the mask misses faint
counterimages, the agent lowers the S/N detection threshold or expands the mask;
it can also remove unrelated emission enclosed by the mask.

\item \textbf{Stage M0: Source grid and global regularization.} With the Stage-A
mass model fixed, the pipeline calibrates the overall regularization strength
for the pixelized source. The agent expands the source grid if reconstructed
emission approaches its boundary.

\item \textbf{Stage M1: Refined isothermal fit.} The pipeline refines the
SIE+shear mass model using the pixelized source. The agent checks that the
observed images trace back to a compact source---ruling out unphysical solutions
such as a demagnified copy of the arcs, an over-magnified solution, or a
noise-dominated reconstruction
\citep[e.g.,][]{nightingale2018autolens,pearson2021cnnplusparametricfitting}---and
adjusts the sampler settings if convergence is insufficient.

\item \textbf{Stage M2: Brightness-adaptive regularization.} With the Stage-M1
mass model fixed, the pipeline uses the M1 source reconstruction to calibrate
brightness-adaptive regularization. The agent widens the prior range if the
posterior for the adaptive hyperparameter reaches its upper bound.

\item \textbf{Stage M3: Final power-law fit.} With the calibrated regularization
fixed, the pipeline fits the EPL+shear mass model. Applying the same residual,
source-reconstruction, and convergence checks, the agent accepts the final model
if it reproduces the primary lensing morphology and records any remaining
localized residuals.
\end{itemize}

Beyond performing the model fits without human intervention, the pipeline
records all configurations, diagnostic plots, and the rationale for each
acceptance, retry, or stopping decision for optional human inspection and
follow-up review.

\begin{figure*}[tp]
    \centering
    \includegraphics[width=\textwidth]{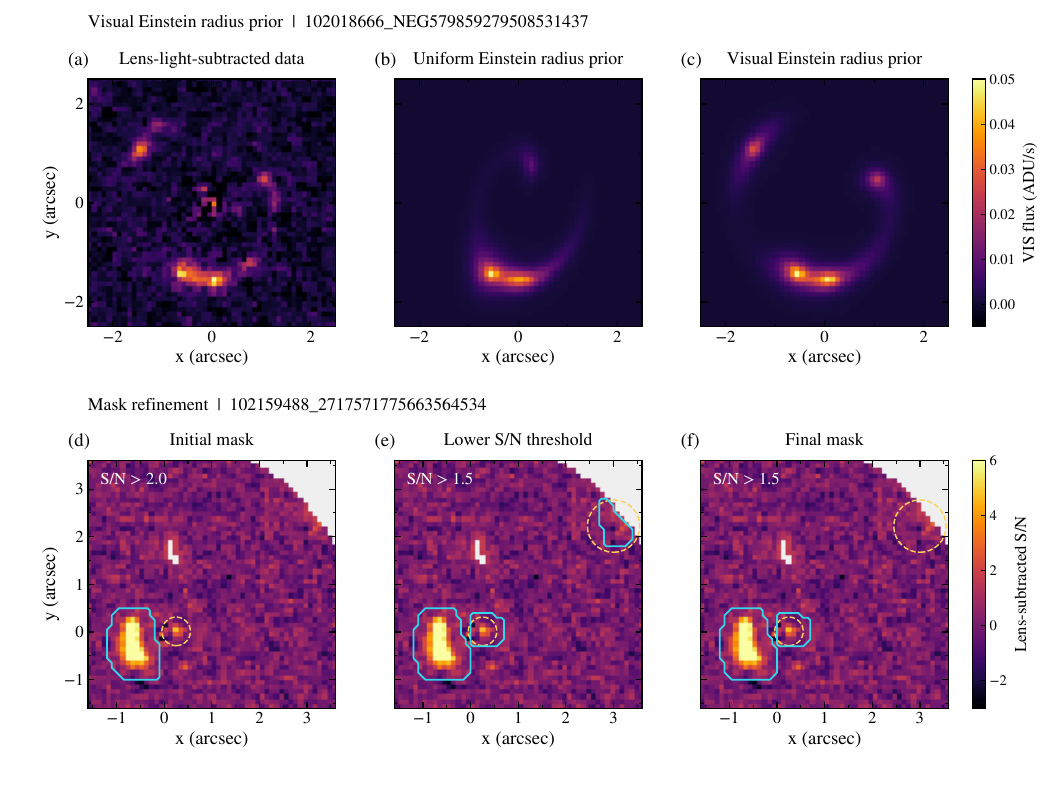}
    \caption{Examples of image-guided agent decisions in two grade-I systems
    (target IDs shown above each row).
    (a--c) Einstein-radius prior comparison in the initial parametric mass and
    MGE light fit (Stage A): (a) lens-light-subtracted
    VIS image, (b) lensed-source model obtained with the default uniform prior
    $\mathcal{U}(0.5^{\prime\prime},3.0^{\prime\prime})$, and (c) model obtained
    with the agent-configured visual prior---an 80\%/20\% mixture of a
    truncated Gaussian ($\mu=1.6^{\prime\prime}$, $\sigma=0.2^{\prime\prime}$)
    and $\mathcal{U}(0.5^{\prime\prime},3.0^{\prime\prime})$---using identical
    imaging, masks, and sampler settings.
    (d--f) Iterative arc-mask refinement (Stage B) on the lens-light-subtracted
    S/N map: (d) the default $\mathrm{S/N}>2.0$ threshold missed the compact
    counterimage; (e) lowering the threshold to $\mathrm{S/N}>1.5$
    captured the counterimage but included an unrelated feature in the upper
    right; and (f) removing that component yielded the final arc mask (cyan
    contours). Dashed circles highlight the counterimage and contaminant
    regions.}
    \label{fig:agent-decisions}
\end{figure*}

\section{Results}\label{sec:results}
\begin{deluxetable}{lcc}[tp]
\tabletypesize{\footnotesize}
\tablewidth{\columnwidth}
\tablecaption{Automated Modeling Results and Human Assessment ($N = 185$)
\label{tab:model-yield}}
\tablehead{
    \colhead{Category} & \colhead{$N$} & \colhead{Fraction (\%)}
}
\startdata
\cutinhead{(a) Modeling Results and Human Quality Grades}
Completed                        & 149 & 80.5 \\
\quad Grade I                    & 126 & 68.1 \\
\quad Grade II                   &  14 &  7.6 \\
\quad Grade III                  &   9 &  4.9 \\
Usable models (Grades I + II)    & 140 & 75.7 \\
Rejected for human review        &  36 & 19.5 \\
\cutinhead{(b) Agent Decision vs.\ Human Assessment}
\multicolumn{1}{l}{Agent Decision} &
\multicolumn{1}{c}{Human: Accept} &
\multicolumn{1}{c}{Human: Reject} \\
Accepted                         & 140 &  9 \\
Rejected for human review        &   0 & 36
\enddata
\tablecomments{Fractions in panel~(a) are relative to the parent sample of
185 \Euclid\ Q1 lenses with published mass models. Quality grades are
assigned through independent human visual inspection of the 149
completed models: Grade~I denotes noise-level image residuals; Grade~II
denotes visible residuals but a well-recovered multiple-image
configuration with usable global lens parameters; and Grade~III
denotes an unacceptable fit. Panel~(b) cross-tabulates the agent
decision against independent human evaluation.
``Human: Accept'' comprises Grades~I and II ($140/149 = 94.0\%$),
whereas ``Human: Reject'' comprises Grade~III false acceptances
($9/149 = 6.0\%$) and the 36 systems rejected for human review (all
confirmed by human inspection).}
\end{deluxetable}

\subsection{Model results and human assessment}\label{sec:model-yield}
Applied to the 185 lenses without human intervention, our automated pipeline
completed final EPL+shear fits for 149 systems ($80.5\%$) and flagged the
remaining 36 ($19.5\%$) for human review (Table~\ref{tab:model-yield}).
Independent human evaluation of the 149 completed models yielded 126 grade-I fits
($68.1\%$ of the full sample), which reconstruct the lensed arcs to the noise
level and constrain both the Einstein radius and the power-law density slope,
and 14 grade-II fits ($7.6\%$), which leave localized residuals that may bias
the density slope but still recover the global multiple-image geometry and
Einstein radius. Together, grades I and II provide 140 usable lens models
($94.0\%$ of the completed fits, or $75.7\%$ of the full sample), leaving only
nine unsuccessful fits ($6.0\%$, grade III). For all 36 systems that the
agents halted and flagged for human review, this human evaluation confirmed
that the automated fits had indeed failed.

Across the 45 systems without an acceptable final model (36 halted and 9
graded III), the automated diagnostics intercepted $80.0\%$ of the problematic
cases before final delivery. Most failures arose either from initial contaminant
masking in crowded fields---where genuine lensed features were masked or
contaminant outskirts were left unmasked (25 cases)---or from faint,
near-noise counterimages that prevented a reliable initial multiple-image fit or
arc mask (13 cases). The remaining 7 cases involved prominent deflector
disks whose light-subtraction residuals blended with the arcs (3 cases),
double-source-plane lenses \citep[2 cases;][]{euclidli2026euclidq1slded},
ambiguous multiple-image identification (1 case), or over-smoothing of the
reconstructed pixelized source (1 case).

Figure~\ref{fig:inspection-reconstruction} illustrates the multi-stage modeling
for a representative grade-I lens
(\texttt{102019595\_NEG630246532498667881}). During initial inspection, the
agents masked five neighboring contaminants and estimated an Einstein radius
of $\simeq 0.8^{\prime\prime}$ to initialize the parametric mass and MGE light
fit (Stage A).
Subtracting the fitted MGE lens-light model cleanly isolated the two lensed arc
regions used to construct the arc mask (Stage B), and the final EPL+shear fit
with a pixelized source (Stage M3) reproduced the lensed arcs to the noise
level while reconstructing a compact source inside the caustic.

\begin{figure*}[tp]
    \centering
    \includegraphics[width=\textwidth]{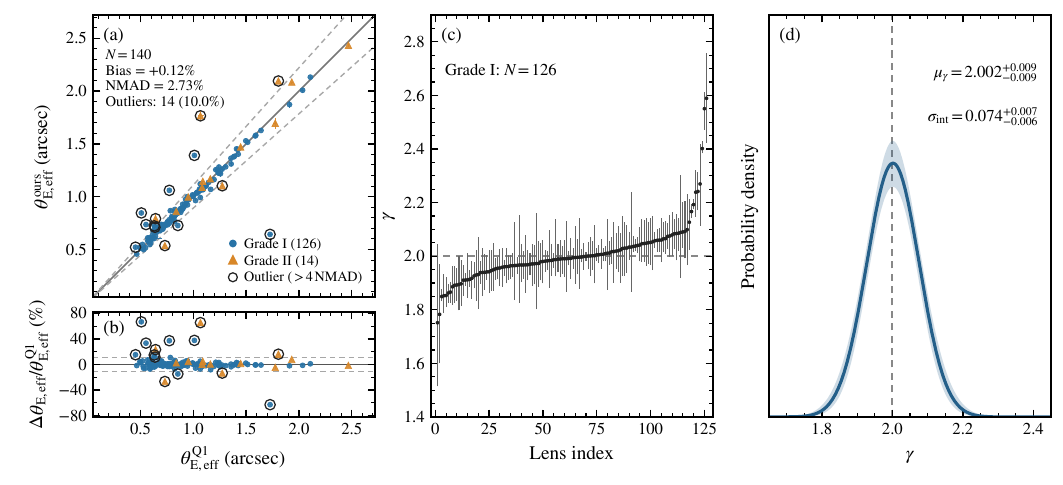}
    \caption{Effective Einstein radii and EPL mass-density slopes inferred by
    our pipeline.
    (a) Effective Einstein radii $\theta_{\rm E,eff}$ for the 140 usable
    models---126 grade I (blue circles) and 14 grade II (orange
    triangles)---compared with the released Q1 SIE+shear models. Points and
    error bars show posterior medians and 68\% credible intervals; the solid
    line marks the one-to-one relation.
    (b) Fractional difference
    $\Delta\theta_{\rm E,eff}/\theta_{\rm E,eff}^{\rm Q1} =
    (\theta_{\rm E,eff}^{\rm ours}-\theta_{\rm E,eff}^{\rm Q1})/
    \theta_{\rm E,eff}^{\rm Q1}$. The solid horizontal line marks the median
    offset, and dashed lines in (a) and (b) indicate $\pm4\,\mathrm{NMAD}$
    around the median offset, with open circles highlighting outliers beyond
    this threshold.
    (c) Individual EPL mass-density slopes $\gamma$ (posterior medians and
    68\% credible intervals, ordered by median slope) for all 126 grade-I
    models.
    (d) Inferred intrinsic Gaussian distribution of $\gamma$ for the grade-I
    lens population, characterized by the mean slope $\mu_\gamma$ and intrinsic
    scatter $\sigma_{\rm int}$ (solid curve and shaded 68\% credible region).
    Dashed gray lines in (c) and (d) mark the isothermal slope $\gamma=2$.}
    \label{fig:einstein-radius-comparison}
\end{figure*}

\subsection{Image-guided modeling decisions}\label{sec:agent-decisions}
Figure~\ref{fig:agent-decisions} illustrates how the agents replicate two
routine tasks of human lens modelers: visually estimating the Einstein radius to
initialize the mass model and iteratively refining the arc mask before pixelized
source reconstruction. For the grade-I lens
\texttt{102018666\_NEG579859279508531437}
(Figures~\ref{fig:agent-decisions}a--c), we compared two initial parametric mass
and MGE light fits (Stage A) that shared identical imaging, contaminant masks,
and sampler settings, differing only in the Einstein-radius prior. Under the broad
default uniform prior $\mathcal{U}(0.5^{\prime\prime},3.0^{\prime\prime})$, the
sampler became trapped in a local maximum that fit the bright lower arc but
missed the upper-left counterimage and underfit the lensed image on the right
(Figure~\ref{fig:agent-decisions}b). When instead guided by the visual
Einstein-radius prior estimated by the inspection agents from the observed
lensed-image separation, the fit recovered the full four-image configuration
(Figure~\ref{fig:agent-decisions}c), providing a reliable starting model for the
subsequent pixelized-source stages.

Figures~\ref{fig:agent-decisions}(d)--(f) show how the Stage-B agent
iteratively inspects and adjusts the arc mask for another grade-I system,
\texttt{102159488\_2717571775663564534}, much as a human lens modeler would.
With the default $\mathrm{S/N}>2.0$ threshold
(Figure~\ref{fig:agent-decisions}d), the compact counterimage had only
four pixels above threshold and was removed by the five-pixel minimum-area cut.
Inspecting the resulting mask, the agent recognized the missing
counterimage and lowered the detection threshold to $\mathrm{S/N}>1.5$. This
adjustment recovered the counterimage, but the lower threshold
simultaneously included an unrelated emission feature in the upper right
(Figure~\ref{fig:agent-decisions}e). After reviewing the updated mask, the agent
removed that feature and accepted the final arc mask
(Figure~\ref{fig:agent-decisions}f) for the subsequent pixelized-source fits
(Stages M0--M3).

\subsection{Einstein radii and mass-density slopes}\label{sec:reconstruction-mass-scale}
To evaluate the inferred lens mass parameters across the sample, we first
compared the effective Einstein radii of all 140 usable models (126 grade I and
14 grade II) with the publicly released Q1 SIE+shear models
(Figures~\ref{fig:einstein-radius-comparison}a and b). To avoid differences in
model-specific mass normalizations, we defined the effective Einstein radius as
$\theta_{\rm E,eff}=\sqrt{A_{\rm t}/\pi}$, where $A_{\rm t}$ is the area
enclosed by the tangential critical curve (including external shear), and
computed $\theta_{\rm E,eff}$ from the posterior samples of our final EPL+shear
fits (Stage M3).

Across all 140 systems, the median
fractional difference relative to Q1 is $+0.12\%$ with a normalized median
absolute deviation (NMAD) of $2.73\%$, demonstrating close overall agreement in
the inferred Einstein radii between our EPL+shear pipeline and the Q1 SIE+shear
models. Fourteen outliers ($10.0\%$), including several grade-I lenses, differ
from the Q1 results by more than $4\,\mathrm{NMAD}$, as expected given the
different modeling choices between the two pipelines---most notably the mass
profile (EPL versus fixed SIE), pixelized source reconstruction, contaminant
masking, and lens-light subtraction.

Whereas the released Q1 SIE models fix the logarithmic mass-density slope to
the isothermal value ($\gamma\equiv 2$, where $\rho\propto r^{-\gamma}$;
\citealt{euclidwalmsley2026euclidq1sldea,euclidbusillo2026euclidq1lemon}), our
final EPL+shear fits (Stage M3) constrain $\gamma$ directly from the extended
lensed images. From a hierarchical Bayesian inference across all 126 grade-I lenses
assuming $\gamma_i\sim\mathcal{N}(\mu_\gamma,\sigma_{\rm int}^2)$
(Appendix~\ref{app:density-slope}), we measure a population mean slope of
$\mu_\gamma = 2.002 \pm 0.009$ and an intrinsic scatter of
$\sigma_{\rm int} = 0.074^{+0.007}_{-0.006}$
(Figures~\ref{fig:einstein-radius-comparison}c and d). This near-isothermal population mean is consistent with previous strong-lensing
studies of massive early-type galaxies from SLACS, BELLS, SL2S, and LSD, whose
samples are typically centered at $z\sim 0.2\text{--}0.6$ while extending out to
$z\sim 0.8\text{--}1$
\citep[e.g.,][]{TreuKoopmans2004,Koopmans2006SLACSIII,Auger2010SLACSX,Sonnenfeld2013SL2SIV,Li2018BELLS,Etherington2023,Tan2024}.
Because \Euclid\ strong lenses are expected to have a higher typical deflector
redshift of $z\sim 0.6\text{--}0.7$
\citep{Collett2015,Cao2024,euclidwalmsley2026euclidq1sldea}, our 126-lens
imaging-only measurement suggests that a near-isothermal average profile remains
representative in this higher-redshift \Euclid\ population, although confirming
any redshift trend and quantifying PSF systematics will require individual
redshifts and higher-fidelity PSF models (Section~\ref{sec:discussion}).

\subsection{Searching for anomalous systems and low-mass perturbers}\label{sec:anomaly-screening}
Beyond population studies of regular single-deflector lenses, our automated
baseline models also help identify systems that warrant more detailed
follow-up. Lenses with dark matter substructure, multiple lens galaxies
\citep[such as merging pairs;][]{Shu2016Offset}, or multiple lensed sources at
different redshifts cannot be fully captured by a smooth single-deflector,
single-source-plane model, and are either flagged by the agents during modeling
or revealed by residuals in the completed fits.

We tested this approach by searching the residual maps from the 149 completed
EPL+shear fits (Stage M3) for localized features that could arise from low-mass
subhalos or line-of-sight halos \citep[e.g.,][]{Vegetti2010,Nightingale2024,He2025}.
A dedicated agent flagged localized anomalies in the residual maps, and human
visual inspection selected two systems for parametric subhalo modeling. Adding a
perturber increased the log evidence over the smooth model by
$\Delta\ln\mathcal{Z}\simeq 5.5$ and $15.2$ for the two candidates,
respectively. We then tested both with free-form potential corrections
\citep{Vegetti2009,Cao2025,Powell2025}. Neither reconstruction showed the
localized positive convergence clump expected from a compact perturber, leaving
both candidates unconfirmed. Even so, this search illustrates how automated
candidate selection can focus computationally intensive follow-up modeling on a
small subset of survey lenses (see Appendix~\ref{app:perturber-search} for
details).

\section{Discussion}\label{sec:discussion}
Our analysis of 185 \Euclid\ Q1 lenses demonstrates that Lens-Orchestra can
automate the visual inspections, failure diagnoses, and configuration
adjustments required to scale physical forward modeling to large surveys. The
need for this capability is underscored by the \Euclid\ DR1 ECLIPSE campaign
\citep{euclidnightingale2026eclipse_c}: although machine-learning preprocessing
and \texttt{PyAutoLens} scaled automated fitting to 15,072 candidates, the team
still relied on over 20 human inspectors to vet the outputs, as summary
goodness-of-fit statistics alone could not distinguish valid fits from
unphysical source reconstructions or lens-light subtraction failures. \citet{euclidnightingale2026eclipse_c} therefore
highlighted agentic AI---capable of jointly evaluating imaging and diagnostics
to perform targeted remodeling---as key to analyzing $\mathcal{O}(10^5)$
lenses. Lens-Orchestra realizes this paradigm on real \Euclid\ data:
astronomers can launch single-object or batch analyses via natural-language
instructions without manually inspecting intermediate fits or editing scripts,
while retaining uniform physical modeling and full diagnostic records.

Currently, Lens-Orchestra relies on general-purpose frontier models
(\texttt{gpt-5.6-sol} and \texttt{gpt-5.6-luna}) guided by expert prompts,
consuming about two billion tokens across the 185 Q1 systems. A natural next
step toward $\mathcal{O}(10^5)$-lens surveys is to use the growing corpus of
human-verified modeling outputs and diagnostic records to fine-tune smaller,
locally deployable open-weight multimodal LLMs (such as \texttt{Qwen3.8-27B}),
internalizing expert diagnostic criteria into a far more affordable and
scalable local agent team.

Our current automated pipeline assumes a single main deflector (SIE or EPL
plus external shear). Neighboring objects that do not noticeably perturb the
arcs are masked as contaminants, whereas systems with secondary deflectors or
companions that significantly distort the lensed images are referred for human
review. For the 126 single-deflector grade-I lenses, our imaging-only EPL fits
constrain the population-mean mass-density slope
(Figures~\ref{fig:einstein-radius-comparison}c and d;
Appendix~\ref{app:density-slope}), subject to two caveats. First, we adopt the
default \Euclid\ MERge (MER) effective VIS PSF, constructed by propagating input
PSF models through resampling and stacking. Because radial density slopes
derived from extended arcs are
sensitive to subtle PSF inaccuracies---which can also induce localized
high-S/N residuals---assessing the accuracy of imaging-only EPL slopes will
require dedicated tests with higher-fidelity PSF models
\citep{euclidnightingale2026eclipse_c}. Second, individual redshifts are not
yet available for these lenses, precluding a detailed study of how $\gamma$
evolves with redshift.

Finally, although all 185 Q1 lenses were modeled autonomously, Lens-Orchestra
also supports an interactive human-in-the-loop mode. Because every stage saves
its intermediate products, settings, and diagnostic records, researchers can
pause an active run---or inspect a completed or halted one---to query the
agent in natural language about its modeling choices, instruct it to revise
settings such as contaminant or arc masks, and resume downstream fits from
that stage. This enables Lens-Orchestra to function both as an automated
survey pipeline and as an interactive co-scientist for complex systems.

\section{Conclusion}\label{sec:conclusion}
We have developed Lens-Orchestra, an automated workflow coupling a team of
multimodal LLM agents to multi-stage physical lens modeling in
\texttt{TinyLensGPU}. Launched via natural-language instructions, the agents
replace manual tuning by inspecting imaging and intermediate diagnostics to
refine priors, masks, and modeling settings without human intervention. Across
185 \Euclid\ Q1 grade-A lens candidates with published mass models, the workflow
produced 140 usable EPL+shear fits ($75.7\%$), including 126 grade-I models
($68.1\%$) fitted to the noise level, while all 36 halted runs were confirmed by
human inspection. The inferred effective Einstein radii agree closely with
released Q1 models (median offset $+0.12\%$, NMAD $2.73\%$), and hierarchical
inference on the 126 grade-I imaging-only fits yields a mean density slope of
$2.002 \pm 0.009$ with an intrinsic scatter of $0.074^{+0.007}_{-0.006}$,
consistent with a near-isothermal average profile. In addition, a dedicated
residual-screening agent guided the selection of two candidate low-mass
perturbers for parametric and free-form substructure modeling; although neither
was confirmed, this test demonstrates how automated screening can focus
computationally intensive follow-up analyses on a small subset of survey
lenses. Coupling
multimodal agents to explicit forward modeling thus offers a scalable path for
uniform physical analysis across $\mathcal{O}(10^4\text{--}10^5)$ survey lenses,
from macro mass profiles to small-scale dark matter structure.

\begin{acknowledgments}
This work was supported by the National Natural Science Foundation of China
(grant Nos.\ 12588202, 12303006, 11773032, and 12022306), the CAS Project
for Young Scientists in Basic Research (grant No.\ YSBR-062), and the science
research grants from the China Manned Space Program (Nos.\ CMS-CSST-2025-A03
and CMS-CSST-2025-A20). We acknowledge the
use of Google Gemini to assist with language editing of this manuscript.
\end{acknowledgments}

\software{
  \href{https://github.com/caoxiaoyue/lens-orchestra}{Lens-Orchestra},
  TinyLensGPU \citep{Cao2025TinyLens},
  \href{https://github.com/openai/codex}{OpenAI Codex CLI},
  nautilus \citep{Lange2023Nautilus},
  JAX \citep{Bradbury2018JAX},
  caskade \citep{Stone2025caskade},
  \href{https://github.com/astroskylee/STCI}{STCI},
  Astropy \citep{Astropy2022},
  Photutils \citep{Bradley2026Photutils},
  scikit-image \citep{VanDerWalt2014ScikitImage},
  NumPy \citep{Harris2020NumPy},
  SciPy \citep{Virtanen2020SciPy},
  Matplotlib \citep{Hunter2007Matplotlib}
}

\restartappendixnumbering
\appendix

\ifSubfilesClassLoaded{\restartappendixnumbering\appendix}{}

\section{Physical Lens Models and Inference}
\label{app:physical-modeling}

\subsection{Forward image modeling and mass parameterization}
For a set of nonlinear lens parameters $\boldsymbol\eta$, the lens equation maps
an image-plane angular coordinate $\boldsymbol\theta$ to the source plane via
$\boldsymbol\beta(\boldsymbol\theta)=\boldsymbol\theta-\boldsymbol\alpha(\boldsymbol\theta;\boldsymbol\eta)$,
where $\boldsymbol\alpha$ is the deflection angle. Because gravitational lensing
conserves surface brightness, the model image is given by
\begin{equation}
    \boldsymbol m=\mathcal P\!\left[
    I_{\rm lens}(\boldsymbol\theta)+
    I_{\rm src}\!\left(\boldsymbol\beta(\boldsymbol\theta)\right)\right],
\end{equation}
where $I_{\rm lens}$ and $I_{\rm src}$ denote the deflector and source surface
brightness distributions, and the linear operator $\mathcal P$ combines pixel
binning and PSF convolution: it averages the surface brightness over a
$4\times4$ subpixel grid within each image pixel and convolves the resulting
pixelized image with the PSF via fast Fourier transforms. Within the unmasked
modeling region $\mathcal M$ (restricted to a circular aperture of radius
$4^{\prime\prime}$ around the deflector center), we assume uncorrelated Gaussian
pixel noise with variance $C_{ii}=\sigma_i^2$ given by the RMS noise map,
yielding
$\chi^2=(\boldsymbol d-\boldsymbol m)^{\mathsf T}
C^{-1}(\boldsymbol d-\boldsymbol m)$ for the observed image $\boldsymbol d$.

We model the surface mass density of the main deflector with an elliptical
power-law (EPL) profile \citep{TessoreMetcalf2015}. In coordinates $(x',y')$
centered on the deflector and aligned with its principal axes, the convergence is
\begin{equation}
    \kappa(R_\epsilon)=\frac{3-\gamma}{2}
    \left(\frac{\theta_{\rm E}}{R_\epsilon}\right)^{\gamma-1},
    \quad R_\epsilon=\sqrt{q x'^2+y'^2/q},
    \label{eq:epl-convergence}
\end{equation}
where $q\in(0,1]$ is the minor-to-major axis ratio, $\gamma$ is the
three-dimensional logarithmic mass-density slope ($\rho\propto r^{-\gamma}$),
and $\theta_{\rm E}$ is the Einstein radius of the EPL profile (whereas
$\theta_{\rm E,eff}$ in Section~\ref{sec:reconstruction-mass-scale} is defined
from the area enclosed by the tangential critical curve of the combined
EPL+shear model). For $\gamma=2$, the EPL profile
reduces to the singular isothermal ellipsoid \citep[SIE;][]{Kormann1994}, and we
compute the EPL deflection angles using the iterative complex-plane series of
\citet{TessoreMetcalf2015}. Because the position angle $\phi$ is periodic over
$[0,\pi)$ and becomes undefined in the circular limit ($q\to1$), we sample the
axis ratio and orientation through the Cartesian ellipticity components
$(e_1,e_2)=\frac{1-q}{1+q}(\cos2\phi,\sin2\phi)$. Tidal perturbations from
nearby galaxies and line-of-sight structures are modeled as an external shear
with lensing potential
\begin{equation}
    \psi_{\rm ext}(x,y)=\tfrac12\gamma_{{\rm ext},1}(x^2-y^2)
    +\gamma_{{\rm ext},2}xy.
\end{equation}

\subsection{Masking, parametric light profiles, and source-plane pixelization}
Prior to model fitting, unrelated neighboring sources within the
$4^{\prime\prime}$ aperture are masked using a segmentation map of the
sky-background-subtracted image, constructed by detecting and deblending
connected regions of at least five pixels above $3\sigma_{\rm sky}$ (where
$\sigma_{\rm sky}$ is the sigma-clipped sky RMS); segments identified as
contaminants during the initial visual inspection are excluded from all modeling
stages.

In the initial parametric fit (Stage A), both the deflector and source surface
brightness distributions are represented using multi-Gaussian expansions
\citep[MGE;][]{He2024MGE}. We model the deflector light with two sets of ten
concentric, co-aligned elliptical Gaussians whose widths $\sigma$ are
logarithmically spaced between $0.01^{\prime\prime}$ and $4.0^{\prime\prime}$,
with each set assigned its own center and ellipticity components $(e_1,e_2)$ to
accommodate isophotal twists, ellipticity gradients, and mild centroid offsets
between the inner and outer regions. The source galaxy is modeled with a single
set of ten concentric, co-aligned elliptical Gaussians with widths spanning
$0.01^{\prime\prime}$ to $1.0^{\prime\prime}$. For each trial set of nonlinear
mass and light geometries, the 30 Gaussian flux amplitudes are solved linearly
via non-negative least squares. Once Stage A is completed, the best-fit
deflector light model is subtracted from the VIS image and held fixed in all
subsequent stages.

Following lens-light subtraction, Stage B constructs an arc mask to isolate the
image-plane region containing the lensed arcs and counterimages, which serves as
the data region reconstructed in the pixelized-source fits (Stages M0--M3). By
default, connected regions of at least five pixels with $\mathrm{S/N}>2$ in the
lens-light-subtracted signal-to-noise (S/N) map are identified and expanded via
morphological closing ($3\times3$ pixels) and dilation ($5\times5$ pixels) to
enclose the faint arc wings. Within this mask, the source surface brightness is
reconstructed on a regular Cartesian grid centered on the Stage-A source position
\citep{WarrenDye2003,Suyu2006}, where the surface brightness at each ray-traced
subpixel position $\boldsymbol\beta(\boldsymbol\theta)$ is evaluated by bilinear
interpolation over the surrounding source-grid pixels. The grid uses a fixed
pixel scale of $0.02^{\prime\prime}$ and defaults to $80\times80$ pixels
($\pm0.8^{\prime\prime}$). If the reconstructed source flux reaches the grid
boundary in Stage M0, the grid is expanded to $100\times100$
($\pm1.0^{\prime\prime}$) or $120\times120$ ($\pm1.2^{\prime\prime}$).

\subsection{Bayesian semi-linear source inversion}
Let $F(\boldsymbol\eta)$ denote the lensing and PSF-convolution operator that
linearly maps the source surface-brightness vector $\boldsymbol s$ to the
lens-light-subtracted lensed-image data $\boldsymbol d'$. With regularization
matrix $H$ and strength $\lambda>0$, minimizing
$Q(\boldsymbol s)=(\boldsymbol d'-F\boldsymbol s)^{\mathsf T}
C^{-1}(\boldsymbol d'-F\boldsymbol s)
+\lambda\boldsymbol s^{\mathsf T}H\boldsymbol s$
with respect to $\boldsymbol s$ gives the most probable source reconstruction
$\widehat{\boldsymbol s}$ \citep{WarrenDye2003,Suyu2006},
\begin{equation}
    (F^{\mathsf T}C^{-1}F+\lambda H)\widehat{\boldsymbol s}
    =F^{\mathsf T}C^{-1}\boldsymbol d',
    \label{eq:normal-equations}
\end{equation}
which we solve in \texttt{TinyLensGPU} \citep{Cao2025TinyLens} using
matrix-free preconditioned conjugate gradients \citep[PCG;][]{Powell2021}.

We adopt a first-order finite-difference (gradient) regularization operator $H$
on the square source grid. Spatially uniform regularization is used during the
initial regularization calibration (Stage M0) and the pixelized-source
SIE+shear fit (Stage M1). For the subsequent regularization re-calibration
(Stage M2) and final EPL+shear fit (Stage M3), we weight the local smoothness
penalty using the Stage-M1 source reconstruction $\boldsymbol s^{(1)}$ so that
bright source structure is less strongly smoothed than the surrounding
background. Each source pixel $i$ is assigned a weight
\begin{equation}
    w_i=\exp\!\left[a(1-u_i)\right],
\end{equation}
where $u_i\in[0,1]$ is the reconstructed source surface brightness
$\boldsymbol s^{(1)}$ normalized by its peak value (with negative values set
to zero), the finite difference between neighboring pixels $i$ and $j$
in $H$ is weighted by $\sqrt{w_i w_j}$, and $a\geq0$ controls the
regularization contrast between the faint background and bright source
regions ($a=0$ recovers uniform regularization).

Following \citet{Suyu2006}, the lens mass parameters and regularization
hyperparameters are inferred by sampling the Bayesian log-evidence,
\begin{equation}
\begin{split}
    \ln\mathcal E={}&
    -\tfrac12\chi^2
    -\tfrac12\lambda\widehat{\boldsymbol s}^{\mathsf T}
    H\widehat{\boldsymbol s}
    -\tfrac12\ln\det(F^{\mathsf T}C^{-1}F+\lambda H)\\
    &+\tfrac12\ln\det(\lambda H)
    -\tfrac12\ln\det(2\pi C).
    \label{eq:approx-source-evidence}
\end{split}
\end{equation}
Because the matrix-free solver does not assemble the full matrix
$F^{\mathsf T}C^{-1}F+\lambda H$ explicitly, we follow \citet{Powell2021} and
approximate the log-determinant terms in
Equation~\eqref{eq:approx-source-evidence} using the PCG preconditioner, which
is constructed as a block-diagonal matrix over $10\times10$-pixel source
sub-blocks. Tests on simulated \Euclid\ lens data show that this approximation
has a negligible impact on the inferred lens mass parameters and source
reconstructions while substantially accelerating the lens modeling.

\subsection{Priors and posterior sampling}
In Stage A, when the visual-inspection agents provide a confident visual
estimate of the Einstein radius, we adopt a mixture prior on $\theta_{\rm E}$
combining an $80\%$ Gaussian component centered on the visual estimate (with a
standard deviation equal to $20\%$ of the estimated value, up to a maximum of
$0.2^{\prime\prime}$) and a $20\%$ uniform component over
$[0.5^{\prime\prime},3.0^{\prime\prime}]$; otherwise, we use a uniform prior
over $[0.5^{\prime\prime},3.0^{\prime\prime}]$. The remaining Stage-A
parameters use Gaussian priors centered on the image center for the deflector
mass and light centers ($\sigma=0.1^{\prime\prime}$) and the source center
($\sigma=0.5^{\prime\prime}$), zero-centered Gaussian priors for the mass and
light ellipticity components ($\sigma=0.2$ and $0.3$, within $[-1,1]$), and
uniform priors over $[-0.2,0.2]$ for $\gamma_{{\rm ext},1}$ and
$\gamma_{{\rm ext},2}$.

Following standard multi-stage lens-modeling practice
\citep[e.g.,][]{nightingale2018autolens,etherington2022nolensleftbehind}, Stages
M1 and M3 place Gaussian priors on $\theta_{\rm E}$, $(e_1,e_2)$, and
$(\gamma_{{\rm ext},1},\gamma_{{\rm ext},2})$ centered on the preceding posterior
medians (Stage A for M1; Stage M1 for M3), with widths set to five times the
corresponding posterior uncertainties (with minimum widths of $10\%$ in
$\theta_{\rm E}$, $0.2$ in $e_1$ and $e_2$, and $0.05$ in
$\gamma_{{\rm ext},1}$ and $\gamma_{{\rm ext},2}$).
The deflector mass center is fixed to the Stage-A posterior median, and Stage M3
adopts a uniform prior over $[1,3]$ for the density slope $\gamma$. For the
source regularization, Stage M0 finds the initial strength
$\lambda_{\rm M0}\in[10^{-10},10^{10}]$ that maximizes the Bayesian evidence;
Stage M1 samples $\ln\lambda$ around $\ln\lambda_{\rm M0}$ with a Gaussian prior
($\sigma=0.15$); and Stage M2 calibrates the adaptive regularization using a
log-uniform prior on $\lambda\in[10^{-6},10^{6}]$ and a uniform prior on
$a\in[0,a_{\max}]$ ($a_{\max}=8$ by default, increased up to $100$ if the
posterior approaches the upper bound). Finally, to prevent unphysical solutions
in which the multiple lensed images do not trace back to a single source
\citep{nightingale2018autolens}, we map the Stage-A predicted multiple-image
positions $\{\boldsymbol\theta_k\}$ to the source plane and apply an exponential
log-likelihood penalty whenever their separation exceeds $0.1^{\prime\prime}$.

We sample the parameter posteriors with \texttt{nautilus}
\citep{Lange2023Nautilus} using $200\text{--}300$ live points. In the automated
workflow, Stage B permits up to ten mask-refinement iterations, and each
physical modeling stage permits up to five refits within a total budget of
fifteen refits per system.

\ifSubfilesClassLoaded{%
    \bibliographystyle{aasjournalv7.1}%
    \bibliography{references}%
}{}
\end{document}

\ifSubfilesClassLoaded{%
    \restartappendixnumbering\appendix\setcounter{section}{1}%
}{}

\section{Hierarchical Inference of the EPL Density Slope}
\label{app:density-slope}

Following \citet[Section~3.2]{Koopmans2006SLACSIII}, we model the intrinsic
three-dimensional logarithmic mass-density slopes $\gamma_i$ of the 126 grade-I
lenses (Section~\ref{sec:reconstruction-mass-scale};
Figures~\ref{fig:einstein-radius-comparison}c and d) as a Gaussian population,
\begin{equation}
    \gamma_i\mid\mu_\gamma,\sigma_{\rm int}
    \sim\mathcal{N}(\mu_\gamma,\sigma_{\rm int}^2),
    \label{eq:slope-population}
\end{equation}
characterized by a population mean $\mu_\gamma$ and intrinsic scatter
$\sigma_{\rm int}$. Rather than approximating each lens constraint as a Gaussian
around a point estimate, we use the full posterior distributions from the final
EPL+shear fits (Stage M3) to account for any asymmetry or non-Gaussian tails.
For lens $i$ with imaging data $D_i$, the individual fit adopts a uniform slope
prior $\pi_i(\gamma)=\mathcal{U}(1,3)$ and yields weighted posterior samples
$(\gamma_{ik},w_{ik})$ with $\sum_k w_{ik}=1$. Dividing out the interim
individual prior to recover the marginal likelihood $p(D_i\mid\gamma)\propto
p(\gamma\mid D_i)/\pi_i(\gamma)$, the likelihood contribution of lens $i$ to the
population hyperparameters is
\begin{equation}
\begin{split}
    \mathcal{L}_i(\mu_\gamma,\sigma_{\rm int})
    &\propto\int p(\gamma\mid D_i)
    \frac{\mathcal{N}(\gamma\mid\mu_\gamma,\sigma_{\rm int}^2)}
    {\pi_i(\gamma)}\,d\gamma\\
    &\simeq\sum_k w_{ik}
    \frac{\mathcal{N}(\gamma_{ik}\mid\mu_\gamma,\sigma_{\rm int}^2)}
    {\pi_i(\gamma_{ik})}.
\end{split}
    \label{eq:slope-recycling}
\end{equation}
Combining the likelihoods of all 126 systems under uniform hyperpriors
$\mu_\gamma\sim\mathcal{U}(1,3)$ and $\sigma_{\rm int}\sim\mathcal{U}(0,1)$
yields $\mu_\gamma = 2.002 \pm 0.009$ and
$\sigma_{\rm int} = 0.074^{+0.007}_{-0.006}$ (posterior medians and
68\% credible intervals). Adopting a conventional Gaussian approximation
for the individual lens likelihoods gives consistent constraints of
$(\mu_\gamma,\sigma_{\rm int})=(2.007,0.078)$, and replacing the uniform prior
on $\sigma_{\rm int}$ with a half-normal prior leaves the inferred parameters
unchanged within the quoted uncertainties.

\ifSubfilesClassLoaded{%
    \bibliographystyle{aasjournalv7.1}%
    \bibliography{references}%
}{}
\end{document}

\ifSubfilesClassLoaded{%
    \restartappendixnumbering\appendix\setcounter{section}{2}%
}{}

\section{Agent-assisted Candidate Selection and Follow-up Searches for Low-mass Perturbers}
\label{app:perturber-search}

Searching for low-mass perturbers by directly fitting subhalo models to every
lens in a large sample is computationally expensive. Instead, we select
candidates from our 149 completed smooth EPL+shear fits (Stage M3) in two steps.
First, a multimodal LLM agent inspects the normalized residual maps to flag
localized anomalies along the lensed arcs that could arise from a small-scale
mass perturbation, while noting whether the features might instead reflect
unmodeled source structure, PSF mismatches, or lens-light subtraction residuals.
The agent flagged three high-priority perturber candidates. Based on subsequent
human visual inspection of these three systems, we selected two plausible
candidates for detailed follow-up: \texttt{102021499\_NEG630283151478334717}
and \texttt{102044823\_NEG519633334274259646}.

We analyze both candidates by combining parametric subhalo modeling with
non-parametric potential corrections. First, we add a spherical pseudo-Jaffe
perturber at the deflector redshift \citep{Vegetti2010}, with convergence
$\kappa_{\rm sub}(R) = (b_{\rm sub}/2)[R^{-1} - (R^2 + r_{\rm t}^2)^{-1/2}]$,
where $R$ is the projected angular distance from the perturber center
$(x_{\rm sub},y_{\rm sub})$, $b_{\rm sub}$ is the angular mass normalization,
and the tidal radius is coupled to the host Einstein radius via
$r_{\rm t} = \sqrt{b_{\rm sub}\,\theta_{\rm E}}$. Sampling
$(x_{\rm sub},y_{\rm sub})$ uniformly across the arc region under shared
macro-model and source-regularization priors, we perform paired nested-sampling
fits to test whether adding the perturber improves the Bayesian evidence
($\Delta\ln\mathcal{Z}$) over the smooth EPL+shear model. Because parametric
searches suffer from degeneracies between the perturber and unmodeled angular
structure in the main deflector that can cause false-positive detections
\citep{ORiordan2024,Stacey2024,Lange2025,Amvrosiadis2025,Stacey2025}, we further
analyze both systems using non-parametric, free-form potential corrections
\citep{Vegetti2009,Cao2025,Powell2025}. This method jointly reconstructs the
pixelized source and a grid-based perturbation to the smooth lensing potential,
in which a genuine low-mass perturber should appear on the resulting
convergence-correction map ($\delta\kappa$) as a localized positive mass clump.

Applying this combined analysis to our two candidates shows that neither system
yields a reliable perturber detection. In the first candidate
(\texttt{102021499\_NEG630283151478334717}), the parametric fit localizes a
perturber near the upper-right arc at
$(x_{\rm sub},y_{\rm sub})=(-0.05^{+0.03}_{-0.04},+0.50^{+0.03}_{-0.03})^{\prime\prime}$
with $b_{\rm sub}=0.12\pm0.03^{\prime\prime}$, increasing the log-evidence by
$\Delta\ln\mathcal{Z}\simeq 5.5$ (corresponding to $\sim 3.3\sigma$).
In strong-lensing substructure analyses, however, an evidence gain of
$\Delta\ln\mathcal{Z}\sim 5$ is not considered significant once systematic
uncertainties in the source reconstruction and PSF are accounted for
\citep[where $\Delta\ln\mathcal{Z}\gtrsim 10$, or $\sim 5\sigma$, is typically
required; e.g.,][]{Nightingale2024}. In the second candidate
(\texttt{102044823\_NEG519633334274259646}), the parametric fit yields a much
larger evidence increase of $\Delta\ln\mathcal{Z}\simeq 15.2$, with
$b_{\rm sub}=0.17^{+0.02}_{-0.03}\,^{\prime\prime}$ and a primary position mode
near the lower-left arc at $(x_{\rm sub},y_{\rm sub})\simeq(-0.39^{\prime\prime},-0.41^{\prime\prime})$.
However, our smooth models do not include higher-order multipoles in the main
deflector, which can be absorbed by a parametric subhalo and mimic a
high-significance detection. Indeed, our non-parametric potential-correction
reconstructions reveal no localized positive convergence clump in either system.
In particular, for \texttt{102044823\_NEG519633334274259646}, the inferred
$\delta\kappa$ map exhibits large-scale positive and negative corrections along
the arc rather than a compact clump, indicating that the parametric pseudo-Jaffe
fit is likely compensating for missing angular complexity in the main
deflector's mass model.

Although we cannot decisively claim a low-mass perturber detection at the S/N
and angular resolution of this \Euclid\ Q1 sample---where confirming weaker
perturbations may require deeper or higher-resolution follow-up imaging---our
analysis demonstrates a complete, end-to-end pipeline for automated perturber
searches that combines automated smooth lens modeling, AI-agent identification
of residual anomalies, human visual vetting, and joint parametric and
non-parametric perturbative verification. By concentrating
computationally intensive perturbative modeling on the small subset of systems
flagged from the smooth-model residuals, this methodology can be scaled to the
$\mathcal{O}(10^5)$ strong-lens systems expected from \Euclid, \textit{Roman},
and CSST.

\ifSubfilesClassLoaded{%
    \bibliographystyle{aasjournalv7.1}%
    \bibliography{references}%
}{}
\end{document}

\clearpage
\bibliographystyle{aasjournalv7.1}
\bibliography{references}

\end{document}